\documentclass[10pt,letterpaper]{article}
\usepackage[margin=0.75in,top=0.72in,bottom=0.75in]{geometry}
\usepackage[utf8]{inputenc}
\usepackage{times}
\usepackage{microtype}
\usepackage{amsmath}
\usepackage{amssymb}
\usepackage{graphicx}
\usepackage{booktabs}
\usepackage[font=small,labelfont=bf]{caption}
\usepackage{array}
\usepackage{tabularx}
\usepackage{multirow}
\usepackage{xcolor}
\definecolor{alethiacite}{HTML}{001472}
\usepackage{textcomp}
\usepackage[round,authoryear]{natbib}
\let\cite\citep
\usepackage{url}
\usepackage{xurl}
\usepackage[colorlinks=true,linkcolor=black,citecolor=alethiacite,urlcolor=alethiacite]{hyperref}
\usepackage{flafter}
\usepackage{placeins}
\usepackage{float}
\usepackage{needspace}
\newcommand{\papertitle}{Neural Defend Ranks \#1 in Global Audio Deepfake Detection Benchmark -- ARCAS on the Podonos Independent Leaderboard}
\newcommand{\paperauthor}{Sivashankar Selvarajan, Piyush Verma, and Sumit Singh}

\makeatletter
\renewcommand\section{\@startsection{section}{1}{\z@}%
  {-1.5ex \@plus -0.5ex \@minus -.2ex}%
  {0.8ex \@plus .2ex}%
  {\raggedright\normalfont\large\bfseries}}
\renewcommand\subsection{\@startsection{subsection}{2}{\z@}%
  {-1.25ex\@plus -0.4ex \@minus -.2ex}%
  {0.55ex \@plus .15ex}%
  {\raggedright\normalfont\normalsize\bfseries}}
\renewcommand\subsubsection{\@startsection{subsubsection}{3}{\z@}%
  {-1ex\@plus -0.35ex \@minus -.2ex}%
  {0.4ex \@plus .1ex}%
  {\raggedright\normalfont\normalsize\itshape}}
\makeatother
\renewenvironment{abstract}{\section*{\centering Abstract}\small}{\par}
\hypersetup{
  pdftitle={\papertitle},
  pdfauthor={\paperauthor},
  pdfsubject={Neural Defend ARCAS on the Podonos independent benchmark},
  pdfkeywords={AI-generated audio detection, foundation model, audio deepfake detection, speech anti-spoofing, robustness, multilingual, partial fakes, replay, benchmark evaluation}
}
\definecolor{arcasblue}{HTML}{2A78D6}

\begin{document}
\twocolumn[{
\begin{center}
\vspace*{0.1in}
\rule{\textwidth}{0.8pt}\par
\vspace{0.2in}
{\LARGE\bfseries \papertitle\par}
\vspace{0.28in}
\rule{\textwidth}{0.8pt}\par
\vspace{0.2in}
{\normalsize\bfseries Sivashankar Selvarajan\textsuperscript{1}\quad Piyush Verma\textsuperscript{1}\quad Sumit Singh\textsuperscript{1}\par}
\vspace{0.06in}
{\small Technical report \textbullet\ Leaderboard snapshot: Podonos release of 29 October 2026 \textbullet\ \href{https://github.com/podonos/audio-dfd-benchmark}{Benchmark repository} \textbullet\ \href{https://www.podonos.com/blog/audio-deepfake-detection-benchmark-neural-defend-leads}{Leaderboard post}\par}
\end{center}
\vskip 0.12in
}]
\footnotetext[1]{Neural Defend Inc. Correspondence to: Sivashankar Selvarajan \texttt{<siva@neuraldefend.com>}.}

\begin{abstract}
Commercial detectors of AI-generated audio are usually promoted with results on internal test sets that buyers cannot inspect. Independent benchmarks with private labels give a more reliable comparison. We report the results of Neural Defend's ARCAS foundation models on the Podonos audio deepfake detection benchmark, an independent leaderboard that scores commercial detection APIs and open-source detectors on one held-out set of real speech and speech from current text-to-speech and voice-cloning systems, delivered in six file formats, against labels that only Podonos holds. In the release of 29 October 2026, which covers 28 systems from 16 commercial vendors and the leading open models, ARCAS 290M and the smaller ARCAS 90M share first place, ahead of every other commercial system, including Resemble AI, Hiya, Pindrop, Hive, and Reality Defender. Both models make fewer errors than any other system by a statistically clear margin, keep false positives and false negatives in balance at their default threshold, process every submitted file without rejection, and sit on the accuracy--latency frontier of the leaderboard. As for every commercial system added since Podonos retired its API runner, the ARCAS predictions were submitted by Neural Defend and scored by Podonos against its private labels. We describe the benchmark's construction and protocol in full, reproduce the complete leaderboard and its history across five releases, and analyse error balance, statistical separation, speed, and the gap between commercial and open-source detectors. The results agree with the companion evaluation of ARCAS 290M on the complete test sets of ten public benchmarks and establish ARCAS as the most accurate AI-generated audio detector evaluated on this independent benchmark.
\end{abstract}

\section{Introduction}

Voice cloning needs only seconds of reference audio, and human listeners identify current synthetic speech as fake about 69\% of the time \cite{muller2026eroding}. Banks, contact centres, telecom carriers, and platforms therefore buy AI-generated audio detection as a service, but vendors' claims are hard to compare. Most vendors report accuracy on internal test sets, on selected public datasets, or on leaderboard snapshots of different dates, often with different definitions of accuracy. Open-source detectors trained on older academic corpora give little guidance either: when Deepfake-Eval-2024 tested them on audio circulated in 2024, their AUROC fell by about half \cite{deepfakeeval2024}.

Independent benchmarks with private labels address this problem. A third party builds a test set from current generators, keeps the labels private, scores every system with one protocol, and publishes the results for anyone to inspect. The Podonos audio deepfake detection benchmark is a private-label leaderboard of this kind that focuses specifically on commercial detection APIs \cite{podonos2026may,podonos2026repo}. Since May 2026 it has grown from eight to 28 systems in five releases, and it now includes 18 commercial entries from 16 vendors and 11 systems with downloadable weights \cite{podonos2026oct}.

In this report we present the results of Neural Defend's ARCAS foundation models on this benchmark. ARCAS 290M is Neural Defend's production model for detecting AI-generated audio; ARCAS 90M is a smaller and faster model of the same family for latency-sensitive use. In the October 2026 release, both models share first place with 99.93\% accuracy: three errors in 4,524 files, with a false-positive rate of 0.1\% and a false-negative rate of 0.0\% \cite{podonos2026oct}. The next system, deetech.ai, makes 20 errors, and the best results from Resemble AI, Hiya, Pindrop, Hive, and Reality Defender range from 24 to more than a thousand errors on the same held-out test set.

The same model also leads on public data, under far broader conditions. In the companion report, ARCAS 290M is evaluated on the complete official test sets of ten public benchmarks, 5,644,079 recordings in total, and reaches 0.14\% equal error rate (EER) on the ASVspoof~5 evaluation set, below every commercial detector reported on that set \cite{arcas290m2026benchmark}. Together the two reports give consistent evidence from a private-label commercial comparison and from complete public test sets.

Our contributions are:
\begin{itemize}
\item \textbf{A complete account of the Podonos benchmark.} We document its test set, telephony tracks, submission protocol, metrics, and provenance rules, collected from every public release, so the results can be read without consulting scattered sources.
\item \textbf{Leading results on an independent benchmark.} ARCAS 290M and ARCAS 90M rank first of 28 systems, ahead of all 15 other commercial vendors, with balanced errors and full coverage.
\item \textbf{Analysis beyond the leaderboard.} We add confidence intervals and significance tests, error-balance and accuracy--latency analysis, the history of the leaderboard across five releases, and a comparison of commercial and open-source detectors.
\item \textbf{Consistency with public benchmarks.} We relate the Podonos results to the full-scale public evaluation of ARCAS 290M and state what the benchmark does and does not measure.
\end{itemize}

\section{Background}

\subsection{How detection vendors report performance}

Published evidence for commercial detectors falls into three groups. The first is internal evaluation: vendors report accuracy on proprietary test sets whose generators, sample counts, and class balance are not released, so the numbers cannot be reproduced or compared across vendors. The second is results on selected public datasets. A few vendors have published peer-reviewed work, such as Reality Defender's ASVspoof~5 challenge system, which reached 5.5\% EER on the Track~1 evaluation set \cite{zhu2024learnfromreal}, but most public numbers come from older corpora or are combined across datasets with vendor-chosen weights. The third is independent leaderboards. The Speech DF Arena scores open and proprietary systems on 14 public datasets, including ASVspoof~5 \cite{dowerah2026dfarena}, and the Podonos benchmark scores commercial APIs on a private-label set of current voice-cloning audio \cite{podonos2026oct}.

Only the third group lets a buyer compare vendors on the same files with the same rules. Leaderboard results still have to be read with their date and release, because rankings change as systems are added: in the five Podonos releases to date, the first place has changed hands three times. Independent results also show what internal tests cannot: how a system behaves at its shipped threshold, how many files it declines, and how its errors split between missed fakes and false alarms. These are the properties a contact centre, bank, or platform experiences once a detector is deployed in production.

\subsection{Why open-source detectors fail on current audio}

Most open-source detectors are trained on ASVspoof 2019, whose attacks were built with text-to-speech and voice-conversion systems from 2018 and earlier. Their accuracy falls sharply on current generators and recording conditions \cite{deepfakeeval2024,mlitw2026,speechfake2025}. On the Podonos benchmark, nine of the 11 downloadable systems score between 47.63\% and 62.89\% accuracy, close to chance on a balanced test set, and one of them labels every clip as fake \cite{podonos2026oct}. Their training data predates the zero-shot voice-cloning systems that attackers use, so their errors reflect outdated training rather than the difficulty of the test set. This gap is the reason commercial detection exists as a product category, and it is why buyers need independent comparisons between commercial systems rather than against academic baselines.

\section{The Podonos Benchmark}

\looseness=-1 This section describes the benchmark as published by Podonos in its repository and in the five release posts \cite{podonos2026repo,podonos2026may,podonos2026jun,podonos2026aug,podonos2026sep,podonos2026oct}. Where the sources differ, we follow the repository.

\subsection{Test set}

\looseness=-1 The studio test set contains 4,524 English clips, balanced between 2,262 real and 2,262 AI-generated recordings, about six hours of audio (Table~\ref{tab:podonos_set}). Real speech comes from three public corpora: VCTK \cite{yamagishi2019vctk} (110 English speakers with a range of accents), LJ Speech \cite{ito2017ljspeech} (one speaker), and LibriTTS train-clean-360 \cite{zen2019libritts} (904 speakers in the full corpus). AI-generated speech comes from about 25 text-to-speech and voice-cloning systems, including commercial APIs such as ElevenLabs and open-weight models such as F5-TTS and Chatterbox; the full list is not published, so detectors cannot be tuned to it.

\looseness=-1 Synthetic clips that fail a Whisper transcription check against their input text are discarded, so the set contains only intelligible speech. Each clip is then converted to one of six file formats, in close to equal numbers. Detectors must therefore handle lossy and lossless compression, as with typical user-uploaded audio.

\begin{table}[H]
\centering
\caption{Composition of the Podonos studio test set \cite{podonos2026repo}. Per-source and per-generator counts are not published.}
\label{tab:podonos_set}
\footnotesize
\setlength{\tabcolsep}{4pt}
\begin{tabular}{@{}lp{5.75cm}@{}}
\toprule
Property & Value \\
\midrule
Clips & 4,524 (2,262 real, 2,262 AI-generated) \\
Language & English \\
Duration & mean 4.80 s; about 6 h in total \\
Real speech & VCTK, LJ Speech, LibriTTS train-clean-360 \\
AI-generated speech & about 25 TTS and voice-cloning systems \\
Quality check & Whisper transcription against input text \\
File formats & MP3 773, WAV 724, FLAC 762, \\
 & OGG 757, M4A 749, WebM 759 \\
Labels & held privately by Podonos \\
\bottomrule
\end{tabular}
\end{table}

\subsection{Telephony tracks}

\looseness=-1 Podonos has also released two telephony versions of the same 4,524 clips, with the same hidden labels (Table~\ref{tab:podonos_tel}). The narrowband track resamples each clip to 8~kHz, applies a 300--3400~Hz band-pass filter, and passes it through one of five codecs used in landline, 2G/3G, and VoIP calls. The wideband track resamples to 16~kHz, applies a 50--7000~Hz filter, and uses one of four codecs used in 4G and 5G calls. Codec assignment is kept private. These tracks reflect the condition most relevant to contact centres, where incoming calls arrive band-limited and compressed by network codecs. No telephony results have been published yet.

\begin{table}[!t]
\centering
\caption{Telephony tracks of the Podonos benchmark: codec, bitrate, number of files, mean PESQ, and mean Whisper word error rate against the clean transcript \cite{podonos2026repo}.}
\label{tab:podonos_tel}
\small
\setlength{\tabcolsep}{4pt}
\begin{tabular}{@{}llrrr@{}}
\toprule
Codec & Bitrate & Files & PESQ & WER \\
\midrule
\multicolumn{5}{@{}l}{\emph{Narrowband, 8 kHz}} \\
G.711 $\mu$-law & 64 kbit/s & 905 & 4.44 & 5.3\% \\
G.711 A-law & 64 kbit/s & 903 & 4.44 & 4.4\% \\
AMR-NB & 12.2 kbit/s & 904 & 4.07 & 8.4\% \\
G.729 & 8 kbit/s & 906 & 3.71 & 6.4\% \\
GSM-FR & 13 kbit/s & 906 & 3.51 & 11.0\% \\
\midrule
\multicolumn{5}{@{}l}{\emph{Wideband, 16 kHz}} \\
EVS-WB & 24.4 kbit/s & 1,129 & 4.05 & 2.0\% \\
AMR-WB & 23.85 kbit/s & 1,133 & 3.72 & 3.2\% \\
EVS-WB & 13.2 kbit/s & 1,130 & 3.69 & 4.0\% \\
AMR-WB & 12.65 kbit/s & 1,132 & 3.27 & 4.0\% \\
\bottomrule
\end{tabular}
\end{table}

\subsection{Protocol and metrics}

\looseness=-1 Each participant submits one prediction file per track, labelling every clip \emph{real} or \emph{fake}, optionally with the latency of each call. Systems are scored at their default threshold, with no per-system tuning. Podonos computes accuracy, $F_1$, false-positive rate (FPR, real speech flagged as fake), and false-negative rate (FNR, fake speech missed). Equal error rate is not reported, because deployed systems use a fixed threshold. Declined or failed files are excluded and reported as a rejection rate.

Latency is the mean per-file inference time in milliseconds, and the real-time factor (RTF) is the mean ratio of latency to clip duration. Podonos measured latency itself, including network round-trips, for the three commercial systems it ran against vendor APIs (Resemble DETECT-3B Omni, Hive, and Reality Defender); for other commercial systems, including ARCAS, latency is reported by the vendor on its own hardware. Since that API runner was retired, every new commercial result, including ARCAS, comes from a vendor-submitted prediction file scored against the private labels.

\needspace{5\baselineskip}
Since October 2026 the rules allow one submission per vendor per month, with one model per track and no scoring of multiple checkpoints or thresholds; timing data are required, and each vendor sees its result before deciding whether to publish it \cite{podonos2026oct}.

\subsection{Systems}\label{sec:systems}

The October 2026 release lists 28 systems \cite{podonos2026oct,podonos2026repo}. Together they cover voice-security specialists, synthetic-media detection platforms, telecom providers, and research laboratories, reflecting the range of detectors buyers can choose from. Eighteen are commercial entries from 16 vendors: Neural Defend ARCAS 290M and ARCAS 90M \cite{neuraldefend}, deetech.ai \cite{deetech}, Resemble DETECT-World and DETECT-3B Omni \cite{resemble}, Hiya \cite{hiya}, Fennura \cite{fennura}, Raid AI \cite{raidai}, Aurigin AI \cite{aurigin}, Whispeak \cite{whispeak}, Momenta \cite{momenta}, Pella Research pellav2 \cite{pella}, Pindrop \cite{pindrop}, DetectifAI \cite{detectifai}, NII Synthetiq Audio v0.8-Beta \cite{niisynthetiq}, Corsound AI \cite{corsound}, Hive \cite{hive}, and Reality Defender \cite{realitydefender,zhu2024learnfromreal}.

Eleven systems have downloadable weights: pellav2, counted in both groups, and ten open-source detectors run locally by Podonos (Section~\ref{sec:opensource}). Each leaderboard row also records how its latency was obtained, separating results Podonos measured directly from those vendors submitted.

\section{Results}

All results in this section come from the leaderboard release of 29 October 2026 \cite{podonos2026oct}; later releases may change the ranking. Confidence intervals, significance tests, and error counts are our calculations from the published accuracy and number of scored files.

\begin{table*}[!htb]
\centering
\caption{Podonos audio deepfake detection leaderboard, October 2026 release \cite{podonos2026oct,podonos2026repo}. Type: A = Neural Defend ARCAS, C = other commercial system, O = open-source model with downloadable weights (pellav2 is both). Rows with $N<4{,}524$ are scored only on the files the system accepted. Latency is the mean time per file; RTF is the mean ratio of latency to clip duration. Latency is measured by Podonos (P) for systems it ran against vendor APIs and reported by the vendor (V) otherwise. References for every system are given in Section~\ref{sec:systems}.}
\label{tab:leaderboard}
\footnotesize
\setlength{\tabcolsep}{3.5pt}
\begin{tabular}{@{}rllrrrrrrrrc@{}}
\toprule
\# & System & Type & $N$ & Rej. (\%) & Acc. (\%) & $F_1$ & FPR (\%) & FNR (\%) & Latency (ms) & RTF & Timing \\
\midrule
\textbf{1} & \textbf{Neural Defend ARCAS 290M} & \textbf{A} & \textbf{4,524} & \textbf{0.0} & \textbf{99.93} & \textbf{0.999} & \textbf{0.1} & \textbf{0.0} & \textbf{118} & \textbf{0.041} & \textbf{V} \\
\textbf{2} & \textbf{Neural Defend ARCAS 90M} & \textbf{A} & \textbf{4,524} & \textbf{0.0} & \textbf{99.93} & \textbf{0.999} & \textbf{0.1} & \textbf{0.0} & \textbf{89} & \textbf{0.03} & \textbf{V} \\
3 & deetech.ai & C & 4,524 & 0.0 & 99.56 & 0.996 & 0.4 & 0.4 & 50 & 0.019 & V \\
4 & Resemble DETECT-World & C & 4,524 & 0.0 & 99.47 & 0.995 & 0.7 & 0.4 & 399 & 0.12 & V \\
5 & Hiya & C & 4,420 & 2.3 & 99.14 & 0.991 & 0.5 & 1.2 & 31 & 0.011 & V \\
6 & Fennura & C & 4,524 & 0.0 & 98.63 & 0.986 & 1.2 & 1.5 & 553 & 0.14 & V \\
7 & Raid AI & C & 4,524 & 0.0 & 98.54 & 0.985 & 1.9 & 1.0 & 93 & 0.033 & V \\
8 & Aurigin AI & C & 4,524 & 0.0 & 98.21 & 0.982 & 2.4 & 1.1 & --- & --- & --- \\
9 & Resemble DETECT-3B Omni & C & 4,524 & 0.0 & 98.05 & 0.981 & 2.5 & 1.4 & 1,164 & 0.4 & P \\
10 & Whispeak & C & 4,524 & 0.0 & 97.70 & 0.977 & 2.9 & 1.7 & 1,052 & 0.39 & V \\
11 & Momenta & C & 4,524 & 0.0 & 97.46 & 0.975 & 2.3 & 2.8 & 42 & 0.011 & V \\
12 & Pella Research (pellav2) & C/O & 4,524 & 0.0 & 95.82 & 0.959 & 5.5 & 2.8 & 57 & 0.021 & V \\
13 & Pindrop & C & 4,524 & 0.0 & 95.05 & 0.951 & 6.2 & 3.7 & 282 & 0.076 & V \\
14 & DetectifAI & C & 4,524 & 0.0 & 94.47 & 0.946 & 8.4 & 2.7 & 24 & 0.0084 & V \\
15 & NII Synthetiq Audio v0.8 & C & 4,524 & 0.0 & 89.57 & 0.892 & 6.6 & 14.2 & 91 & 0.024 & V \\
16 & Corsound AI & C & 3,875 & 14.3 & 87.79 & 0.865 & 1.0 & 23.1 & 180 & 0.035 & V \\
17 & Hive & C & 4,524 & 0.0 & 83.53 & 0.808 & 2.4 & 30.5 & 881 & 0.34 & P \\
18 & Reality Defender & C & 3,745 & 17.2 & 71.27 & 0.770 & 53.7 & 3.6 & 5,718 & 1.52 & P \\
\midrule
19 & NII AntiDeepfake & O & 4,524 & 0.0 & 70.47 & 0.584 & 0.4 & 58.6 & 91 & 0.024 & P \\
20 & Wav2Vec2 (ASVspoof 2019 LA) & O & 4,524 & 0.0 & 62.89 & 0.514 & 13.4 & 60.8 & 622 & 0.14 & P \\
21 & AST (ASVspoof 5) & O & 4,524 & 0.0 & 56.83 & 0.657 & 69.0 & 17.4 & 5 & 0.0017 & P \\
22 & Wav2Vec2 (2024 mix) & O & 4,524 & 0.0 & 55.55 & 0.499 & 33.1 & 55.8 & 219 & 0.056 & P \\
23 & Deepfake-V2 (W2V2-base) & O & 4,524 & 0.0 & 53.03 & 0.162 & 3.1 & 90.9 & 94 & 0.027 & P \\
24 & AST (VoxCelebSpoof) & O & 4,524 & 0.0 & 50.99 & 0.048 & 0.5 & 97.5 & 8 & 0.003 & P \\
25 & RawNet2 (ASVspoof 2019 LA) & O & 4,524 & 0.0 & 50.66 & 0.430 & 35.9 & 62.7 & 94 & 0.035 & P \\
26 & LCNN-LFCC (ASVspoof 2019 LA) & O & 4,524 & 0.0 & 50.00 & 0.667 & 100.0 & 0.0 & 23 & 0.0056 & P \\
27 & AASIST (ASVspoof 2019 LA) & O & 4,524 & 0.0 & 48.17 & 0.486 & 52.6 & 51.1 & 322 & 0.11 & P \\
28 & AASIST3 (ASVspoof 5) & O & 4,524 & 0.0 & 47.63 & 0.029 & 6.3 & 98.4 & 363 & 0.13 & P \\
\bottomrule
\end{tabular}
\end{table*}

\subsection{Overall ranking}\label{sec:ranking}

ARCAS 290M and ARCAS 90M share first place among 28 systems, each with 99.93\% accuracy, $F_1$ of 0.999, FPR of 0.1\%, and FNR of 0.0\% (Table~\ref{tab:leaderboard}). Both scored all 4,524 files without rejection. Each made three errors; the published rates imply that all three, or two of them, were real recordings flagged as AI-generated, and that at most one AI-generated clip was missed.

\begin{figure}[H]\centering
\includegraphics[width=\columnwidth]{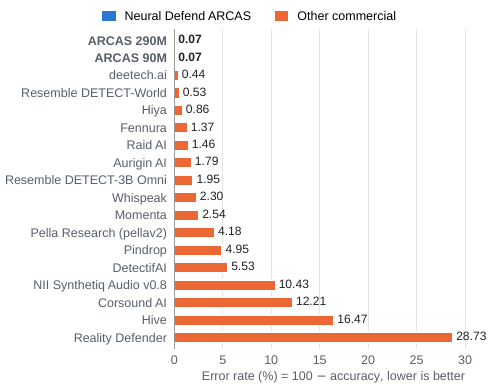}
\caption{Error rate (100 $-$ accuracy, \%) of Neural Defend ARCAS and the 16 other commercial systems on the Podonos benchmark, October 2026. Hiya, Corsound AI, and Reality Defender are scored only on the files they accepted. Source: Podonos \cite{podonos2026oct}.}\label{fig:errors}
\end{figure}

Figure~\ref{fig:errors} shows the error rate of every commercial system. The nearest competitors are deetech.ai (99.56\%, 20 errors), Resemble DETECT-World (99.47\%, 24 errors), and Hiya (99.14\% on the 4,420 files it accepted). ARCAS makes 85\% fewer errors than deetech.ai and 88\% fewer than Resemble DETECT-World. Against established vendors the gap is larger: Resemble DETECT-3B Omni makes 88 errors, Pindrop 224, Hive 745, and Reality Defender 1,076 on the 3,745 files it accepted.

\subsection{Statistical significance}\label{sec:stats}

\begin{figure}[H]\centering
\includegraphics[width=\columnwidth]{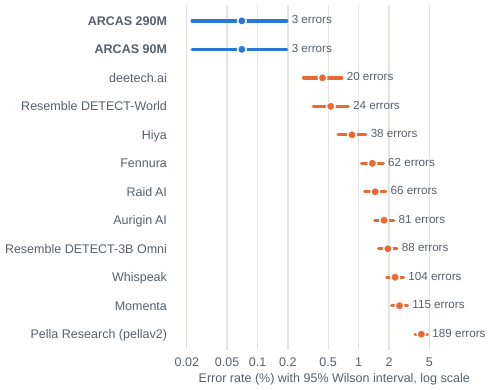}
\caption{Error rate with 95\% Wilson interval for the twelve most accurate commercial systems, computed from the published accuracy and number of scored files. Labels give the implied number of errors.}\label{fig:ci}
\end{figure}

Small accuracy differences can arise by chance, so we compare error rates with 95\% Wilson intervals (Figure~\ref{fig:ci}). The interval for ARCAS, 0.02--0.20\%, overlaps no other system's interval; the closest, deetech.ai, spans 0.29--0.68\%. A one-sided Fisher exact test, treating the errors as independent, gives $p=0.0002$ for three versus 20 errors. A paired test would be more precise but needs per-file predictions, which Podonos does not publish. Only ARCAS has a statistically clear lead over the field; the intervals of each of the next five commercial systems overlap those of their neighbours.

\subsection{False positives and false negatives}\label{sec:balance}

\begin{figure}[H]\centering
\includegraphics[width=\columnwidth]{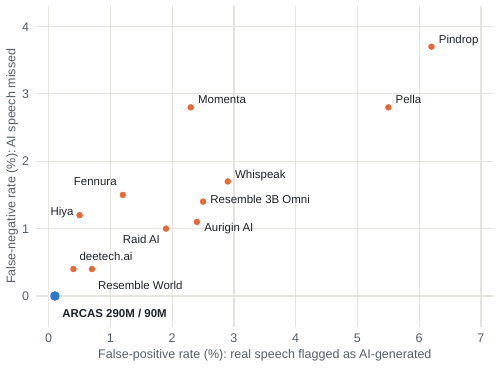}
\caption{False-positive and false-negative rates of commercial systems at their default thresholds; the best possible result is the lower-left corner. ARCAS 290M and ARCAS 90M have identical rates. Off-scale (FPR/FNR, \%): DetectifAI 8.4/2.7, NII Synthetiq 6.6/14.2, Corsound AI 1.0/23.1, Hive 2.4/30.5, Reality Defender 53.7/3.6.}\label{fig:fpr_fnr}
\end{figure}

The Podonos protocol scores each system at its own default threshold, so the balance between false positives and false negatives shows how well the threshold suits real use. ARCAS has the lowest FPR and the lowest FNR of all commercial systems (Figure~\ref{fig:fpr_fnr}). Several other systems keep one error type low at the cost of the other. Corsound AI flags only 1.0\% of real speech but misses 23.1\% of AI-generated speech, and Hive misses 30.5\%. Reality Defender misses 3.6\% of AI-generated speech but flags 53.7\% of real speech as fake. For a contact centre, the first pattern lets fraud through and the second blocks genuine customers; ARCAS does neither.

\subsection{Latency}\label{sec:speed}

\begin{figure}[H]\centering
\includegraphics[width=\columnwidth]{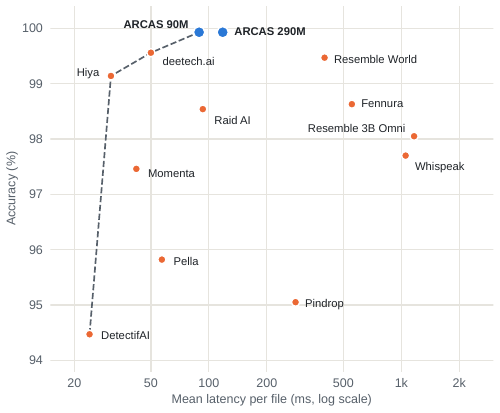}
\caption{Accuracy against mean latency per file for commercial systems with timing data and accuracy of at least 94\%. The dashed line joins systems that no other system beats on both measures. Latency is vendor-reported except for Resemble DETECT-3B Omni, Hive, and Reality Defender.}\label{fig:latency}
\end{figure}

Detection in live calls and content moderation needs low latency. ARCAS 90M reports 89~ms per file (RTF 0.030) and ARCAS 290M 118~ms (RTF 0.041), so both process audio more than 24 times faster than real time. Both are faster than Resemble DETECT-World (399~ms), Pindrop (282~ms), Whispeak (1,052~ms), and Resemble DETECT-3B Omni (1,164~ms). Figure~\ref{fig:latency} plots accuracy against latency. The frontier of systems that no other system beats on both measures runs through DetectifAI, Hiya, deetech.ai, and ARCAS 90M; ARCAS 90M is the most accurate point on it, and the three faster systems make between 6 and 83 times as many errors. Speed also determines coverage in deployment: a detector this fast can screen every call or upload as it arrives, rather than a sample. These latencies are vendor-reported and indicative.

\subsection{Coverage}\label{sec:coverage}

A detector that declines short or unusual files leaves them unprotected. ARCAS scored every file. Hiya rejected 2.3\% of files, Corsound AI 14.3\%, and Reality Defender 17.2\%, including all clips shorter than 1.5~s \cite{podonos2026may}. Their accuracies are measured on the remaining files only; Podonos reports that re-scoring full-coverage systems on those subsets moves their accuracy by $-1.2$ to $+4.2$ percentage points \cite{podonos2026repo}. Full coverage matters in practice because the clips are short, with a mean of 4.80~s, close to the length of a greeting or a single sentence on a call. A detector that skips short audio leaves unchecked exactly the segments an attacker would use. ARCAS returned a verdict for every clip in all six file formats, so its reported accuracy applies to the complete test set.

\subsection{Leaderboard history}\label{sec:progress}

\begin{figure}[H]\centering
\includegraphics[width=\columnwidth]{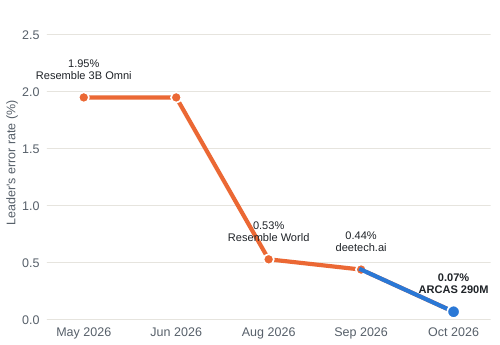}
\caption{Error rate of the leading system in each Podonos release. ARCAS 290M and ARCAS 90M entered in October 2026.}\label{fig:evolution}
\end{figure}

Table~\ref{tab:rounds} summarizes the five releases. The error rate of the leading system fell from 1.95\% in May and June to 0.53\% in August and 0.44\% in September (Figure~\ref{fig:evolution}). ARCAS reduced it to 0.07\% in October, an 84\% reduction relative to the previous leader and a larger step than any earlier release. Over the same period the number of systems above 95\% accuracy grew from two to 13, so ARCAS leads a field that is itself improving quickly.

\begin{table}[H]
\centering
\caption{The five public releases of the Podonos leaderboard in 2026 \cite{podonos2026may,podonos2026jun,podonos2026aug,podonos2026sep,podonos2026oct}. $\geq$95\%: number of systems with at least 95\% accuracy. Leader accuracy in parentheses.}
\label{tab:rounds}
\footnotesize
\setlength{\tabcolsep}{4pt}
\begin{tabular}{@{}lrrl@{}}
\toprule
Release & Systems & $\geq$95\% & Leader (accuracy, \%) \\
\midrule
5 May & 8 & 2 & Resemble DETECT-3B Omni (98.05) \\
23 Jun & 16 & 4 & Resemble DETECT-3B Omni (98.05) \\
14 Aug & 18 & 6 & Resemble DETECT-World (99.47) \\
27 Sep & 23 & 8 & deetech.ai (99.56) \\
29 Oct & 28 & 13 & \textbf{Neural Defend ARCAS (99.93)} \\
\bottomrule
\end{tabular}
\end{table}

\subsection{Commercial and open-source detectors}\label{sec:opensource}

The ten open-source detectors are NII AntiDeepfake \cite{ge2025posttraining,hf_niiantideepfake}; Wav2Vec2 trained on ASVspoof 2019 LA \cite{hf_w2v2_2019}; AST trained on ASVspoof~5 \cite{hf_ast_asv5}; Wav2Vec2 trained on a 2024 data mix \cite{hf_w2v2_2024}; Deepfake-V2, based on wav2vec~2.0 base \cite{hf_deepfakev2}; AST trained on VoxCelebSpoof \cite{hf_ast_vox}; RawNet2 \cite{tak2021rawnet2,hf_rawnet2} and LCNN-LFCC \cite{hf_lcnn}, both trained on ASVspoof 2019 LA; AASIST trained on ASVspoof 2019 LA \cite{jung2022aasist,gh_aasist}; and AASIST3 trained on ASVspoof~5 \cite{hf_aasist3}. Pella Research's pellav2 reaches 95.82\%, and NII AntiDeepfake 70.47\%. The other nine score 47.63--62.89\%. ARCAS makes about one sixtieth as many errors as pellav2, so academic detectors are no substitute for a production system. The gap is not explained by model size alone: several open models use large self-supervised encoders, yet without current training data they cannot separate modern voice clones from real speech. Detection quality depends on keeping pace with the generators that attackers use.

\section{Discussion}

\subsection{Consistency with public benchmarks}

A leaderboard of 4,524 English clips cannot by itself show how a detector behaves across languages, channels, and generator families. The companion report answers that question for ARCAS 290M on the complete official test sets of ten public benchmarks, 5,644,079 recordings in total \cite{arcas290m2026benchmark}. The two sources agree. On ASVspoof~5, ARCAS 290M reaches 0.14\% EER on all 680,774 Track~1 evaluation records, against 0.38--0.79\% for the commercial detectors on the Speech DF Arena leaderboard \cite{dowerah2026dfarena} and 5.5\% for Reality Defender's published challenge system \cite{zhu2024learnfromreal}. Across five public benchmarks of whole-file AI-generated speech it detects 99.60\% of 4.80 million fakes, and on ASVspoof~5 its accuracy stays at or above 99.69\% in all 12 codec and telephone conditions. The Podonos lead is therefore not specific to one test set.

\subsection{What the benchmark measures}

The Podonos benchmark tests the situation most buyers face: short English clips from current voice-cloning systems, delivered in mixed file formats, scored at the vendor's default threshold with no access to the labels. Its private labels, monthly submission limit, and single-model rule make it hard to tune a system to the test set. Its balanced classes make accuracy, FPR, and FNR directly interpretable, and its rejection accounting exposes systems that decline difficult files.

\subsection{Scope and known limitations}

Several limits apply to these results, as they do to every system on the leaderboard. The ARCAS predictions were submitted by Neural Defend and scored by Podonos against its private labels, as for every commercial system added since the API runner was retired; the reported latencies were measured by Neural Defend on its own hardware. The test set is English only and small enough that differences of a few errors are uncertain, which is why we report confidence intervals. Podonos does not publish per-generator or per-format results for ARCAS, and no results have yet been released for the 8~kHz and 16~kHz telephony tracks. Finally, a leaderboard is a snapshot, and later releases may change the ranking.

\section{Conclusion}

On the Podonos independent benchmark, Neural Defend's ARCAS 290M and ARCAS 90M rank first among 28 systems and ahead of all 15 other commercial vendors, with three errors in 4,524 files, the lowest false-positive and false-negative rates of any commercial system, full coverage, and latency below 120~ms per file. Their lead over the next system is the only statistically clear gap among the leading systems, and their entry cut the leading error rate by 84\%. Together with the full-scale public evaluation of ARCAS 290M, these results establish ARCAS as the most accurate AI-generated audio detector on the independent evidence reviewed here. We encourage other vendors to report results on independent private-label benchmarks and on complete public test sets, so that buyers can compare detectors on the same evidence.

\section*{Acknowledgments}
We thank Podonos for building and maintaining the independent audio deepfake detection benchmark, for scoring every submission against its private labels, and for publishing its test-set design, protocol, and complete leaderboard, which made this report possible. We also thank the vendors and researchers whose systems appear on the leaderboard, and the creators of the VCTK \cite{yamagishi2019vctk}, LJ Speech \cite{ito2017ljspeech}, and LibriTTS \cite{zen2019libritts} corpora used to build the test set.

\setlength{\bibsep}{2pt plus 1pt}
\renewcommand{\bibfont}{\small}
\bibliographystyle{plainnat}
\bibliography{references}

\end{document}